\documentclass[epj]{svjour}

\usepackage{graphics}

\begin{document}

\title{Study of $\omega$--meson production in
\emph{pp} collisions at ANKE}

\author{S.~Barsov\inst{1}\and
 M.~B\"uscher\inst{2}\and
 M.~Hartmann\inst{2}\and
 V.~Hejny\inst{2}\and
 A.~Kacharava\inst{3,4}\and
 I.~Keshelashvili\inst{2,4}\and
 A.~Khoukaz\inst{5}\and
 V.~Koptev\inst{1}\and
 P.~Kulessa\inst{6}\and
 A.~Kulikov\inst{7}\and
 I.~Lehmann\inst{8}\and
 V.~Leontyev\inst{2}\and
 G.~Macharashvili\inst{4,7}\and
 Y.~Maeda\inst{9}\and
 T.~Mersmann\inst{5}\and
 S.~Merzliakov\inst{2,7}\and
 S.~Mikirtytchyants\inst{1}\and
 A.~Mussgiller\inst{3}\and
 D.~Oellers\inst{2}\and
 H.~Ohm\inst{2}\and
 F.~Rathmann\inst{2}\and
 R.~Schleichert\inst{2}\thanks{email: r.schleichert@fz-juelich.de}\and
 H.~Seyfarth\inst{2}\thanks{Present address: Merkatorstrasse 7, D-52428 J\"ulich}\and
 H.~Str\"oher\inst{2}\and
 S.~Trusov\inst{10,11}\and
 Y.~Valdau\inst{2,1}\and
 P.~W\"ustner\inst{12}\and
 S.~Yaschenko\inst{3,7}\and
 C.~Wilkin\inst{13}
}

\authorrunning{S.~Barsov \emph{et.al.}}

 \institute{High Energy Department, Petersburg Nuclear Physics Institute, 188350 Gatchina, Russia 
 \and Institut f\"ur Kernphysik, Forschungszentrum J\"ulich, D-52425 J\"ulich, Germany 
 \and Physikalisches Institut II, Universit\"at Erlangen--N\"urnberg, 91058 Erlangen, Germany 
 \and High Energy Physics Institute, Tbilisi State University, 0186 Tbilisi, Georgia 
 \and Institut f\"ur Kernphysik, Universit\"at M\"unster, 48149 M\"unster, Germany 
 \and Institute of Nuclear Physics, 31342 Cracow, Poland 
 \and Laboratory of Nuclear Problems, Joint Institute for Nuclear Research, 141980 Dubna, Russia 
 \and Physics and Astronomy Department, University of Glasgow, Glasgow G12 8QQ, UK 
 \and Research Center for Nuclear Physics, Osaka University, Ibaraki, 567-0047 Osaka, Japan 
 \and Institut f\"ur Kern- und Hadronenphysik, Forschungszentrum Rossendorf, 01314 Dresden, Germany 
 \and Lomonosov Moscow State University Skobeltsyn Institute of Nuclear Physics, 119992 Moscow, Russia 
 \and Zentralinstitut f\"ur Elektronik, Forschungszentrum J\"ulich, 52425 J\"ulich, Germany 
 \and Physics and Astronomy Department, UCL, London, WC1E 6BT, UK 
  }

\date{Received: 17 August 2006 / Revised version:}
\abstract{The production of $\omega$-mesons in the
$pp\to{}pp\omega$ reaction has been investigated with the
COSY--ANKE spectrometer for excess energies of 60 and 92\,MeV by
detecting the two final protons and reconstructing their missing
mass. The large multipion background was subtracted using an
event-by-event transformation of the proton momenta between the
two energies. Differential distributions and total cross sections
were obtained after careful studies of possible systematic
uncertainties in the overall ANKE acceptance. The results are
compared with the predictions of theoretical models. Combined with
data on the $\phi$-meson, a more refined estimate is made of the
Okubo-Zweig-Iizuka rule violation in the $\phi/\omega$ production
ratio.
} 
\maketitle

\section{Introduction}
\label{intro}

The nucleon--nucleon system is strongly coupled to channels that
contain one or multiple mesons. The production of these in
$NN$-collisions, preferably near their respective thresholds, will
test and constrain theoretical models since, in these cases, only
few partial waves contribute. The production of pions and heavier
pseudoscalar mesons including kaons has been systematically
studied in proton-proton interactions in the near--threshold
region~\cite{Hanhart}. Now vector mesons, such as the $\rho$ and
the $\omega$, are significant contributors to the $NN$ force at
short distances but for these particles much less information is
available on their production mechanism due to the lack of
experimental data. The relative production of the isoscalar
$\omega$ and $\phi$ mesons has recently received renewed interest
in connection with the so--called Okubo--Zweig--Iizuka (OZI)
rule~\cite{OZI}. This ratio may give information on the admixture
of strange quarks in the nucleon only if the production mechanism
is known to be the same for both
mesons~\cite{Nakayama,Kampfer,Fuchs}. This can be best controlled
through near--threshold measurements but, as yet, the experimental
data set is rather limited.

In the $\omega$ case, the $pp\to{}pp\omega$ total cross section
was measured for excess energies $Q\leq 30\,$MeV at SATURNE by
detecting the two protons in the SPESIII magnetic spectrometer and
identifying the meson by the missing--mass method~\cite{SPES}.
In a more exclusive experiment, the DISTO collaboration also
measured $\omega$ production but at much higher energy ($Q\approx
320\,$MeV)~\cite{DISTO}. The only data between were taken at
COSY--TOF, a non--magnetic time--of--flight spectrometer, where
the reaction was studied at $Q=92$ and 173\,MeV~\cite{TOF}. This
latter energy is in fact the lowest for which differential
distributions are available. Extra data are therefore required to
fill the gap and confirm the energy dependence. For this purpose
we here present $\omega$ results obtained using the COSY--ANKE
facility at 60 and 92\,MeV. It is relevant within the OZI context
to note that the DISTO group measured $\phi$ production at
82\,MeV, and we have recently published similar data at 18.5,
34.5, and 75.9\,MeV~\cite{Hartmann}.

The experimental facility at our disposal is described in
section~\ref{expset}, where particle identification and efficiency
determination, as well as the extraction of the luminosity, are
also discussed. Since the $\omega$-meson production is identified
from the missing mass of two final protons, one of the major
problems is a large background under the $\omega$ peak arising
from the production of two or more pions. We use the same
kinematic transformation as proposed in the SPESIII analysis to
estimate this multipion background by employing data obtained at
the other energy. This is the crucial element of the
$pp\rightarrow{}pp\omega$ event selection, which is treated in
detail in section~\ref{wselect}. The acceptance of the two protons
from the $pp\to{}pp\omega$ reaction in ANKE was far from complete
and so a thorough investigation of various one--dimensional
differential distributions was undertaken and compared with the
results of simulations obtained on the basis of various model
assumptions to be found in section~\ref{difdist}. The values of
the total cross sections given in section~\ref{xtot} are compared
with the results of several theoretical approaches. Also to be
found there is a discussion of how our data influence the
extraction of the OZI ratio from near--threshold $\omega/\phi$
production. Our conclusions and thoughts for future investigations
are presented in section~\ref{conclude}.

\section{Experimental set-up and raw data analysis}
\label{expset}

The measurement has been carried out at the ANKE
installation~\cite{ANKE}, using the internal hydrogen cluster--jet
target~\cite{Khoukaz} and COSY circulating proton beam. To reduce
systematic effects, the beam momentum was changed every 10 minutes
between 2.85\,GeV/c and 2.95\,GeV/c, corresponding to excess
energies of 60\,MeV and 92\,MeV, respectively. Charged particles
resulting from beam--target interactions, and going forward with
laboratory polar angles up to 20$^{\circ}$, were deflected by the
central dipole D2 onto the ANKE detector systems. In the present
investigation, one final proton was detected in the ANKE forward
detector (FD), passing through a set of three multiwire
proportional chambers (MWPC) and two layers of scintillators
placed downstream of D2 and close to the beam pipe. The other
proton was detected in coincidence in the positive charge detector
system (PD), positioned to the right of D2. This system includes
one layer of 23 thin start scintillators (SA) placed close to the
exit window of the D2 vacuum chamber, two MWPCs, and one layer of
six large stop scintillators (SW) within the ANKE Side Wall. Only
that part of the PD which has acceptance for the $pp \rightarrow
pp \omega$ reaction was used in this experiment.

In order to suppress the very large count rate from proton--proton
elastic scattering, the coincidence of signals from the FD and SW
scintillators was chosen as the main trigger. However, 1$\%$ of
the count rate from the FD scintillators alone was added to the
trigger stream so that the data could be normalised. The momenta
of these particles were reconstructed and $pp$ elastic scattering
identified from the clear peak in the missing--mass distribution
at the nominal proton mass. The data were corrected for the
efficiency of the FD MWPCs on an event--by--event basis, as
described below. The number of protons elastically scattered
between laboratory angles of $6.6^{\circ}$ and $8.8^{\circ}$ was
determined from a fit of the peak by a Gaussian, with a polynomial
background. The contribution of the background under the peak was
found to be $(1.8\pm0.8)\%$. Using predictions for the
differential cross section from the SAID database~\cite{SAID}, the
same average luminosity of $7.5\times
10^{30}\,\textrm{cm}^{-2}\textrm{s}^{-1}$ was found for the two
incident momenta, the values being constant over the angular range
to better than $2\%$. The corresponding integral luminosities were
0.280~pb$^{-1}$ and 0.273~pb$^{-1}$ at 2.85\,GeV/c and
2.95\,GeV/c, respectively. The overall systematic error of about
5$\%$ was estimated taking into account the uncertainty in the
SAID calibration standard ($\sim3\%$), the efficiency
determination ($\sim1\%$), and the $0.1^{\circ}$ precision in the
determination of the scattering angle ($\sim3\%$).
\begin{figure}[h]
\begin{center}
\resizebox{0.43\textwidth}{!}{%
\includegraphics{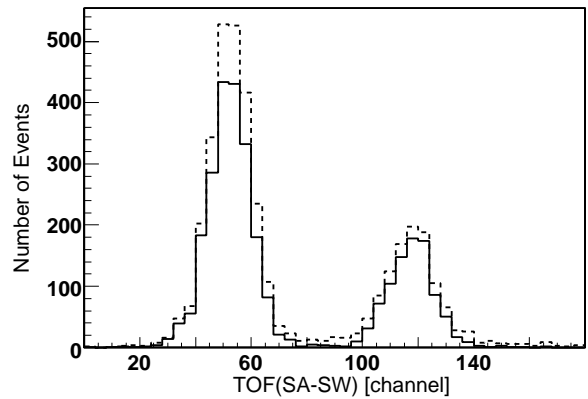}
} \caption{The time--of--flight distributions of particles
detected in one of the SA and one of the SW scintillators,
obtained before (dashed) and after (solid line) the determination
of their momenta. The reduced number of events in the latter case
is the result of the efficiencies of the MWPCs and the
track--finding algorithm. The left and right peaks arise,
respectively, from pions and protons emitted from the target.}
\label{TofPd}
\end{center}
\end{figure}

As the $pp \rightarrow pp \omega$ reaction has to be identified
from the missing mass with respect to the two protons, these were
distinguished from other particles, mostly pions, using the
time--of--flight (TOF) technique and momentum determination.

Although the SW scintillators are capable of delivering a momentum
acceptance of the PD up to 1.6\,GeV/c, in the actual configuration
of the equipment the momentum range had to be limited to be below
1.1\,GeV/c. Under these conditions, the time of flight between SA
and SW scintillators alone provided sufficient particle
identification. As an example of this, Fig.~\ref{TofPd} represents
the TOF spectrum corresponding to the worst case of pion--proton
separation.

The total efficiency of the momentum reconstruction in the PD
could be determined for each of the SA--SW combinations because
both of the MWPCs are placed between these scintillators. The
numbers of protons before and after the reconstruction of their
momenta were extracted from the corresponding TOF spectrum, which
is similar to the one presented in Fig.~\ref{TofPd}. The
accidental background was significantly suppressed by the
requirement to have some particle track reconstructed in the FD.
The efficiency was found to decrease smoothly from $92\%$ to
$86\%$ with increasing particle momentum and track inclination but
to remain constant in the vertical direction for all SA--SW
combinations involved in the later analysis.

\begin{figure}[htb]
\begin{center}
\resizebox{0.43\textwidth}{!}
{\includegraphics{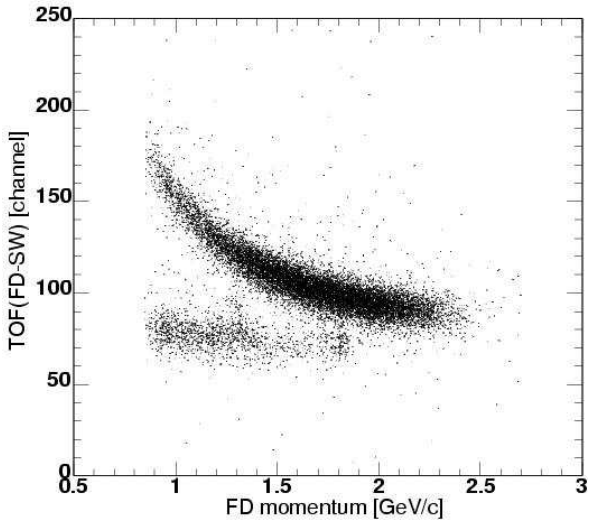}} \caption{The TOF difference
between particles detected in the FD and protons in SW
scintillators versus the momentum of particles in the FD. Protons
detected in the FD produce the upper curved band whereas the TOF
of pions is almost independent of the momentum. } \label{TofvP}
\end{center}
\end{figure}

In contrast to the PD, the two layers of scintillators used in the
FD are placed very close to each other and, in order to identify
protons, the TOF difference between the SW and FD scintillators
had to be combined with the measurement of the momentum of
particles detected in the FD, as shown in Fig.~\ref{TofvP}.
Coincident protons were selected in the PD as described above and
their momenta reconstructed to make a correction for the different
trajectory lengths. In the momentum range of interest, which is
below 1.8\,GeV/c, the proton and pion bands are well separated
except in the vicinity of 1.75\,GeV/c. Here the numbers of pions
are strongly enhanced by the proton--neutron final state
interaction in the $pp \rightarrow p n \pi^+$ reaction. However,
if misidentified as protons, such pions will produce a peak in the
$pp$ missing--mass distribution at 0.65\,GeV/c$^2$ and
0.68\,GeV/c$^2$ for 2.85\,GeV/c and 2.95\,GeV/c, respectively. This
allows the possibility to control the proton selection also in the
overlap region of Fig.~\ref{TofvP}. The total loss of protons as a
result of the selection was estimated to be $(7\pm2)\%$.

To determine the efficiency of a single FD MWPC, the trajectory of
a particle has been found using only information from the other
two MWPCs. Having calculated the point at which the trajectory
intersects the sensitive plane of the MWPC under investigation,
the presence of a valid neighbouring cluster has been checked to
satisfy all requirements applied in the track--finding
algorithm~\cite{Dymov}. This procedure resulted in an efficiency
map for each of the sensitive planes, which was subsequently used
to introduce the efficiency correction on an event--by--event
basis. Only one sensitive plane among the seven involved in the FD
track reconstruction was found to have an inhomogeneous
distribution of efficiency in approximately $30\%$ of its
sensitive area, with an average value of $88\%$. All the others
had homogeneous maps with average efficiencies above $96\%$.

The influence of the total efficiency corrections on the
missing--mass distributions has been checked using the ratio of
corrected to uncorrected spectra, which was found to be constant
to within experimental errors. The average efficiency was about
75\%.

\section{Selection of \boldmath $\mathsf{pp \to pp\omega}$ events}
\label{wselect}

\begin{figure}[htb]
\begin{center}
\resizebox{0.43\textwidth}{!} {\includegraphics{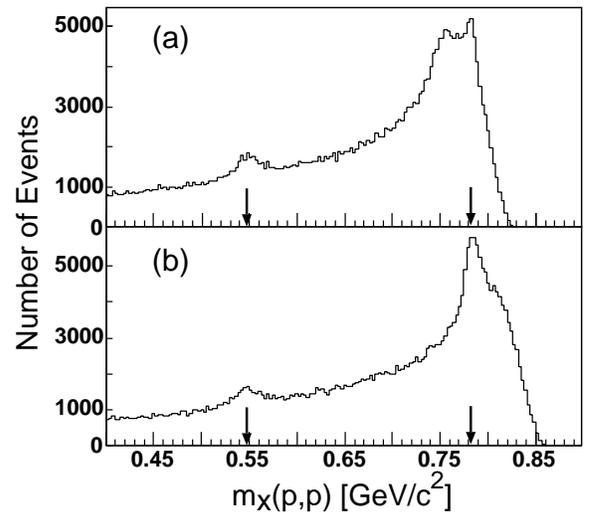}}
\caption{The missing--mass distributions of the $pp \rightarrow pp
X$ reaction measured (a) at 2.85\,GeV/c and (b) at 2.95\,GeV/c.
There is clear evidence for peaks corresponding to the production
of both the $\eta$ and $\omega$ (indicated by arrows) but sitting
on a large multipion background.} \label{separate}
\end{center}
\end{figure}

Histograms representing the overall missing--mass ($m_X$)
distributions from the $pp \rightarrow pp X$ reaction at the two
beam momenta are shown in Fig.~\ref{separate}. 

Both spectra show similar features. 
There is a sharp rise from the kinematic limit
on the right due to the multiparticle phase space, whereas the
more gentle fall to the left is mainly a reflection of the ANKE
acceptance. Sitting on top of this multipion background there are
small peaks due to the production of the $\eta$-- and
$\omega$--mesons. Evidence for the $pp\rightarrow pp \pi^0$
reaction was also found below 0.4\,GeV/c$^2$, but the pion peak
has a large width because the effects of momentum uncertainties
are significantly amplified in this region of large excess
energies.

\begin{figure}[htb]
\begin{center}
\resizebox{0.35\textwidth}{!}
{\includegraphics{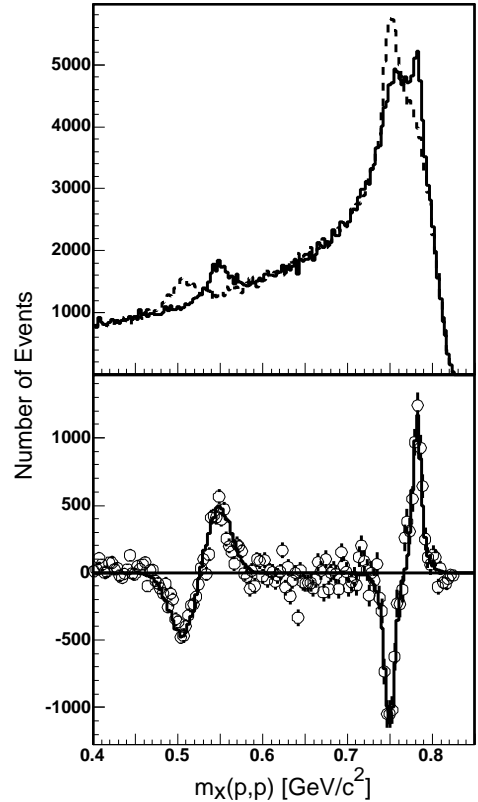}}%
\caption {Upper frame: The missing--mass distribution of the $pp
\rightarrow pp X$ reaction measured at 2.85\,GeV/c (solid line,
Fig.~\ref{separate}a) is shown together with the analogous
distribution obtained at 2.95\,GeV/c (Fig.~\ref{separate}b) and
kinematically transformed to the 2.85\,GeV/c (dashed line), as
described in the text. Lower frame: The bin--by--bin difference of
these two distributions is represented by points. The solid lines
show simulations of the $pp\rightarrow{}pp\eta$ and
$pp\rightarrow{}pp\omega$ reactions convoluted with acceptance.
The widths of the peaks, which are strongly influenced by effects
of momentum reconstruction, are very well reproduced.}
\label{mxtot}
\end{center}
\end{figure}

The accuracy of the missing--mass reconstruction was verified
using the $pp\rightarrow{}pp\eta$ and the $pp\rightarrow{}pn\pi^+$
reactions, with the results being presented in Table~\ref{mxrec}.
\begin{table}[h]
\caption{Missing masses of various reactions in MeV/c$^2$ for the
two beam momenta.} \label{mxrec}
\begin{center}
\begin{tabular}{lll}
\hline\noalign{\smallskip}
$ m_X $ & 2.85\,GeV/c & 2.95\,GeV/c \\
\noalign{\smallskip}\hline\noalign{\smallskip}
$m_n(\pi^+_{PD},p_{F\!D})$& $939.3\pm0.1$ & $939.6\pm 0.1$ \\
$m_n(p_{PD},\pi^+_{F\!D})$ & $939.2\pm 0.2$ & $939.8\pm 0.2$ \\
$m_{\eta} (p_{PD},p_{F\!D})$ & $547.8\pm 0.1$ & $547.4\pm 0.1$ \\
$m_{\omega} (p_{PD}, p_{F\!D})$ & $782.3\pm 0.4$ & $782.7\pm 0.4$ \\
\noalign{\smallskip}\hline
\end{tabular}
\end{center}
\end{table}
In the pion case, the neutron mass was reconstructed in both
possible variants, either when the proton was found in the FD and
the pion in the PD or \emph{vice versa}. Given the good agreement
achieved in these cases, one should expect to find the $\omega$
peak at its nominal position. However, in the vicinity of the $\omega$
mass, the large background from multipion production is strongly
peaked, due to the specific acceptance of the ANKE magnetic
spectrometer, and the challenge is to extract the number of
$pp\rightarrow{}pp\omega$ events.
To select the $\omega$ signal from Fig.~\ref{separate}, a
kinematical transformation of the experimental missing--mass
distribution from one beam momentum to the other was applied, as
illustrated in Fig.~\ref{mxtot}. This approach was proposed and
successfully used in the study of $\omega$ production at lower
excess energies with the SPESIII spectrometer~\cite{SPES}. It
avoids the need to construct specific models for a background that
is unlikely to follow phase space. The scheme involves an
event--by--event Lorentz transformation of the momenta of both
final protons from the laboratory system where they were actually
measured to the laboratory system at another beam momentum. If the
phase space of the multipion production processes is much larger
than the experimental acceptance, the shape of the missing--mass
distribution related to this background is expected to be fixed by
the acceptance and to remain the same in both systems. On the
other hand, a peak corresponding to the production of a single
meson is shifted by approximately the difference in
centre--of--mass energies, as seen clearly for $\eta$ production
in Fig.~\ref{mxtot}. The solid histogram in the top panel of this
figure represents the missing--mass distribution actually measured
at 2.85\,GeV/c (see Fig.~\ref{separate}a). The dashed histogram
represents the result of such a transformation of the
missing--mass distribution obtained at the 2.95\,GeV/c beam
momentum (see Fig.~\ref{separate}b) to the 2.85\,GeV/c. The
transformed distribution is normalised by the ratio of the sums of
events in the 0.6--0.7\,GeV/c$^2$ range, which gives a factor of
$1.054\pm 0.006$.

The difference between the actual and transformed distributions,
shown by points in the lower frame of Fig.~\ref{mxtot},
demonstrates clear positive (negative) peaks corresponding to
$\eta$ and $\omega$ production in the actual (transformed) cases.
These peaks are in a good agreement with the simulations of these
reactions (solid line). The missing--mass distributions simulated
at 2.95\,GeV/c were transformed to 2.85\,GeV/c and
subtracted in the same way as the experimental one. The $\eta$ and
$\omega$ experimental widths arise mainly from the uncertainties
in the momentum reconstruction. Finally, the difference spectrum
for $m_X>0.65\,$MeV/c$^2$ was fitted by two Gaussian
functions in order to extract the number of events in the $\omega$
peak, its statistical uncertainty, and the position of the peak at
$Q=60\,$MeV (see Table~\ref{mxrec}).

The same procedure was applied to extract the total number of
$\omega$-mesons detected at $Q=92$\,MeV. In this case the
missing--mass distribution measured at a beam momentum of 2.85\,GeV/c 
was transformed to 2.95\,GeV/c.

However, the good agreement found for the shapes of the background
spectra in the region between the $\eta$ and $\omega$ peaks does
not necessarily prove that the same is true also underneath the
$\omega$ peaks. In principle, up to six pions might be produced at
our energies, but only those with two and three pions in the
final state have
phase spaces much larger than the ANKE acceptance. Although the total
cross sections for the production of four or more pions are at
least an order of magnitude smaller, they have larger acceptances
in ANKE and populate missing masses mostly near the kinematical
limit.

The direct way of verifying the shape of the background under the
$\omega$ peak, by interpolating between two transformed
distributions measured above and below the given excess energy, is
not possible here since data were taken at only two energies.  An
attempt to describe the multipion background with the help of
phase--space simulations, adjusting the relative contribution of
reactions with different number of pions, was also undertaken.
However, no set
of parameters could simultaneously provide an appropriate
reproduction of the background shape in different parts of the
acceptance, probably due to the neglect of all resonance effects.
We have therefore used the following procedure to evaluate
the possible variation in the shape near the kinematical limit in
the missing mass.

\begin{figure}[htb]
\begin{center}
\resizebox{0.4\textwidth}{!}{%
\includegraphics{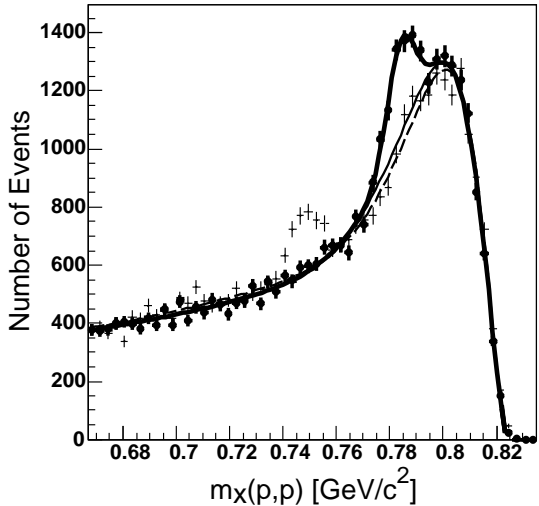}}
\caption {The missing--mass distribution of the
$pp\rightarrow{}ppX$ reaction obtained at 2.85\,GeV/c within a
restricted part of the total acceptance (closed circles) is shown together with the
corresponding distribution kinematically transformed from the
2.95\,GeV/c to 2.85\,GeV/c (crosses). The thick solid line
represents a fit to the distribution at 2.85\,GeV/c while the thin
one shows the background function under the $\omega$ peak found in
this case. The same fit procedure was applied to the transformed
distribution and the resulting background function shown by the
dashed line.} \label{mxratio}
\end{center}
\end{figure}

Due to the characteristics of the ANKE installation, the shape of
background varies across the total spectrometer acceptance. Using
this feature, an acceptance range was found where the shape of
background could be directly established from the fit procedure.
The results from this are shown in Fig.~\ref{mxratio}. Both the
measured and the transformed distributions were then separately
fitted, using the same function to describe the background,
together with a Gaussian function for the $\omega$ peak. The
transformed distribution in Fig.~\ref{mxratio} was scaled by
$1.086\pm 0.011$, \emph{i.e.}\ the ratio of the integrals under
the background functions obtained in this way. The ratio of the
background functions is not quite constant over the range but,
apart from the 5--7 MeV/c$^2$ interval next to the kinematical
limit, the deviation from unity is rather small. The difference
between the background functions observed near the maximum of the
distribution does not exceed $5\%$, changing sign in vicinity of
0.75\,GeV/c$^2$, and staying at the $2\%$ level at smaller masses.
Since the accuracy of the fit is similar to these deviations, a
variation of $3\%$ has been considered as a measure of the
systematics when the number of events under the $\omega$ peak
could be determined only from the difference of experimental
distributions at different energies.

The total numbers of $\omega$-mesons deduced from fits to the
distribution at 2.85\,GeV/c shown in the lower frame of
Fig.~\ref{mxtot}, and the
analogous one at 2.95\,GeV/c, were found to be:\\
\centerline{$5560\pm540(\textrm{stat.})\pm 660(\textrm{syst.})$ at
$Q=60$\,MeV,}
 \centerline{$5360\pm 480(\textrm{stat.})\pm 720(\textrm{syst.})$
at $Q=92$\,MeV.} The large statistical errors are the result of the
subtraction procedure as well as the need to describe two peaks of
opposite sign situated close to each other.

\section{Analysis of differential distributions} \label{difdist}

The restriction on the particle momenta registered in the PD leads
to some mismatch in the FD and the PD momentum ranges and this in
turn results in a limitation on the $pp\rightarrow{}pp\omega$
phase space available within the ANKE acceptance. The acceptance
corrections can therefore not be done in a model--independent way
and some assumptions have to be made on the differential
distributions to determine the total cross section. To test the
validity of these assumptions, we have compared a number of
one--dimensional $pp\to{}pp\omega$ experimental distributions with
the results of simulations. The following observables have been
considered:
\begin{itemize}
\item{The excitation energy in the two--proton rest frame: \\
$\varepsilon_{pp} = M_{\rm inv}(p,p) - 2 m_p $. }%
\item{The excitation energy of the $\omega$-meson and the proton
detected in the PD:
$\varepsilon_{\omega p}=M_{\rm inv}(\omega,p)-m_p-m_{\omega}$.\\
In this case, the total range of the Dalitz plot projection could
be covered due to the large FD momentum acceptance. The range of
the other projection, where the proton was detected in the PD, was
less than 40\,MeV and 75\,MeV at excess energies of 60\,MeV
and 92\,MeV, respectively. }%
\item{The angle of the $\omega$-meson with respect to the beam
direction in the overall centre--of--mass system:
$\cos\Theta_{\rm cm}$.} %
\item{The angle of a proton momentum $\vec q$ with respect to the
beam direction in the $pp$ rest frame: $\cos\Theta_{pp}$.}
\item{The angle of $\vec q$ with respect to the momentum of the
$\omega$-meson in the $pp$ rest frame: $\cos\Psi_{pp}$.}
\end{itemize}

The number of $\omega$-mesons in each bin of such a distribution
was determined from the corresponding missing--mass spectrum using
approaches that depended slightly on the shape of background and
the position of the $\omega$ peak relative to the upper edge of
the spectrum.
\begin{itemize}
\item{When, as in Fig.~\ref{mxratio}, the shape of the background
could be clearly established, a preference was given to the fit
procedure described above. The missing--mass transformation was
used to check that the description of the background below the
$\omega$ peaks was reasonable. It was found that the normalisation
factor of the transformed missing--mass distribution to the
measured one is not constant, but may vary by up to $7\%$.
Therefore, this factor had to be determined for each bin. }%
\item{In cases where the fit procedure could only be applied at a
single excess energy, generally at 92\,MeV, the background
function for the corresponding transformed distribution was
obtained. A simple scaling of this function was then allowed
which, together with a Gaussian $\omega$ peak, was used to fit the
distribution measured at the other excess energy. The scaling
parameter thus obtained was used for the normalisation of the
transformed distribution to the measured one. The number of
$\omega$-mesons obtained as a result of the fit was verified using
the difference of distributions of the type shown in
Fig.~\ref{mxtot}. This procedure was also applied at
$\varepsilon_{pp} > 60\,$MeV, where the distributions transformed
from 2.85\,GeV/c to 2.95\,GeV/c no
longer contained the $\omega$ peak.}%
\item{The scenarios outlined above describe more than $80\%$ of
our data. For the remainder, the transformed distribution was
normalised by the ratio of the numbers determined after excluding
the range covered by the actual and the transformed
$\omega$--peaks from each of the spectra.
It was verified that the sum of background events under the
actual $\omega$--peak found in this and the previous two cases
is equal within errors to the total number of background events
shown in Fig.~\ref{mxtot}. }
\end{itemize}

\begin{figure}[htb!]
\begin{center}
\resizebox{0.48\textwidth}{!}{%
\includegraphics{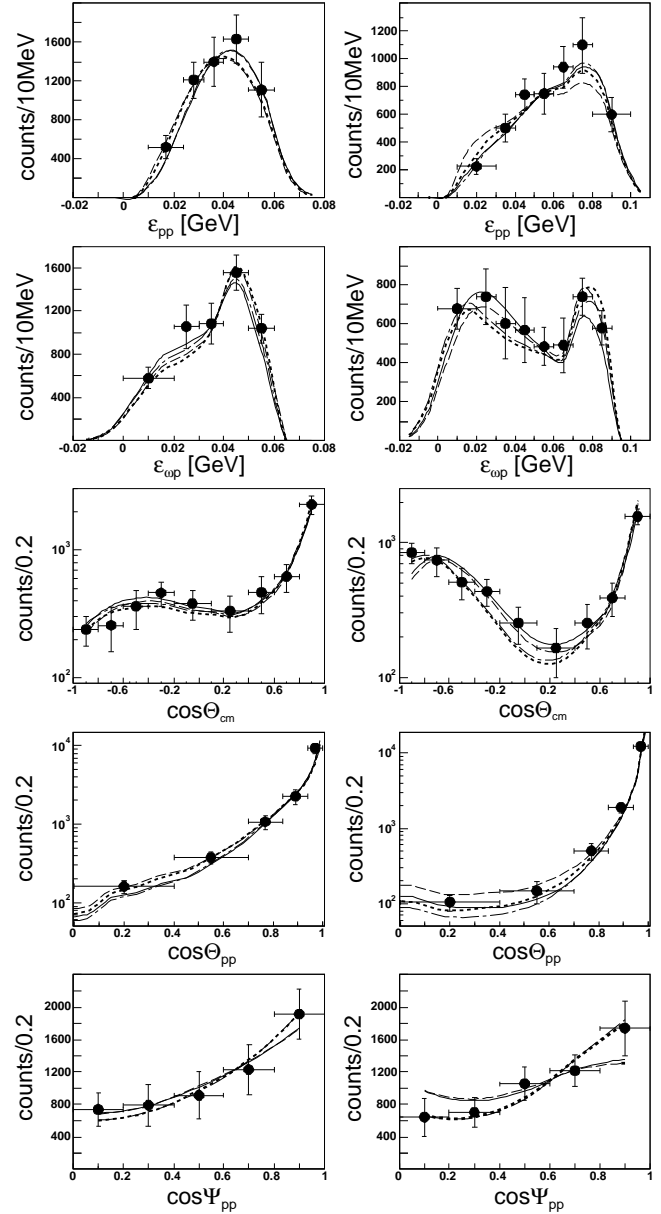}}
\end{center}
\caption {The distributions of experimental (points) and simulated
(lines) $\omega$-meson yields for different reaction observables
at 60\,MeV (left panels) and 92\,MeV (right panels). The different
simulations correspond to pure phase space PS (solid line), full
FSI (dashed line), phase space distorted by an angular dependence
in the $\Psi_{pp}$ variable (dash--dotted line), and partial FSI
with the distorted angular distribution (dotted line), as
described in the text. The corresponding total cross sections are
given in Table~\ref{xvalues}.} \label{dNwdVar}
\end{figure}

In all the cases, different assumptions regarding the background
functions and/or different normalisation factors were tested in
order to establish possible systematic errors, and these are
incorporated into the vertical error bars on the points presented
in Fig.~\ref{dNwdVar}.

The $pp\rightarrow{}pp\omega$ reaction was simulated at both
excess energies using the PLUTO event generator~\cite{PLUTO} under
different assumptions on the distribution of events over the total
phase space of the reaction. Both protons were traced through the
ANKE acceptance with the help of the GEANT3 package~\cite{GEANT}.
To check the tracing procedure, the momentum and angular
acceptances of all the scintillators, as well as their images in
the MWPCs, have been evaluated and compared with the corresponding
experimental ones. They were found to coincide to within the
experimental errors. Finally, the total number of accepted events
simulated at each excess energy was normalised to the
corresponding total number of $\omega$-mesons obtained in the
experiment.

The solid lines in the Fig.~\ref{dNwdVar} demonstrate the
distributions expected in the case of a three--body phase space
(PS), whereas those shown by dashed lines take into account the
final state interaction (FSI) between the two protons. Following
Goldberger and Watson~\cite{Gold}, the FSI enhancement factor $I(q)$
was calculated using the Jost function $J(q)$:
\begin{equation}
\label{FSI}
 I(q) = |J(q)|^{-2} = \left(\frac{q^2 + \beta^2}{q^2 +\alpha^2}\right)\:,
\end{equation}
where $q=|\vec{q}|$ is the magnitude of the proton momentum in the $pp$
rest frame. On the basis of the $pp$ scattering length and effective range
we took $\alpha =-20.5\,$MeV/c, $\beta =167\,$MeV/c.

\begin{figure}[htb]
\begin{center}
\resizebox{0.4\textwidth}{!}{%
 \includegraphics{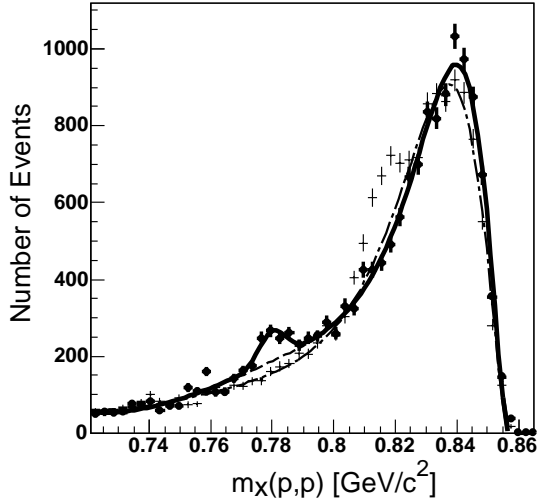}}
\end{center}
\caption {Missing--mass distribution of the $pp\rightarrow{}ppX$
reactions obtained at $Q =92\,$MeV in the $10<\varepsilon_{pp}<
30\,$MeV range (closed circles) is shown together with the
corresponding transformed distribution (crosses). The thick solid
line represents the fit result for the measured spectrum. The
dashed and dashed--dotted lines show two extreme cases of the
background behaviour. The maximum systematic error was established
to be $25\%$. } \label{MxFSI}
\end{figure}

Due to the selection of one proton in the FD and the other in the
PD, the experiment had no acceptance for $\varepsilon_{pp} <
10\,$MeV. As a consequence, and as shown by the left panels of
Fig.~\ref{dNwdVar}, the distributions at $Q=60\,$MeV cannot
distinguish between the pure PS and FSI models. However, at $Q
=92\,$MeV, the experimental yield in the
$10<\varepsilon_{pp}<30\,$MeV range is about a factor of two lower
than that expected on the basis of the FSI model. It must be noted
that the yield in this bin can be determined very reliably, as
shown by the corresponding missing--mass spectrum in
Fig.~\ref{MxFSI}. The fit procedure applied to the missing--mass
distribution at $Q=92\,$MeV would give an even smaller yield than
that found taking into account the behavior of the transformed
spectrum under the peak. Different background functions were
tested, but none of them could increase the yield by more than
$25\%$ assuming a smooth shape under the $\omega$ peak.

The low experimental yield at $Q=92\,$MeV for $\varepsilon_{pp}
<30\,$MeV is similar to that observed in the analogous
$pp\to{}pp\eta$ reaction at 74\,MeV, where the effects of the FSI
in the $pp$ system were found to be very small~\cite{Pauly}. This
may be an indication that the $pp$ $S$--wave is weak and that
higher $pp$ partial waves are important. In this connection it
should be noted that the distributions in the proton angles in the
$pp$ rest frame, especially, $\Psi_{pp}$, would allow some
anisotropy. This contrasts with the yield distributions in the
meson centre--of--mass angle, which correspond quite well to an
isotropic $\textrm{d}\sigma/\textrm{d}\cos\Theta_{\rm cm}$ at both
the excess energies. The TOF collaboration reported a similar
behaviour~\cite{TOF}.

The dash--dotted lines in Fig.~\ref{dNwdVar} were calculated
assuming that the three--body phase space is modified by a factor
\begin{equation}
\label{FACTOR}
1 + a\, P_2(\cos\Psi_{pp})\,, %
\end{equation}
where $P_2(\cos\Psi_{pp})$ is a Legendre polynomial, and $a=0.6$
at $Q =92\,$MeV. It is interesting that this assumption gives a
better description of the data, not only for the $\cos\Psi_{pp}$
distribution, but also for that of $\cos\Theta_{\rm cm}$ in the
$-0.9$ region. At $Q
=60\,$MeV we would expect any anisotropy to be smaller. The case
of $a=0.2$ was tested but, for all the observables other than
$\cos\Psi_{pp}$, the results differ little from the PS case.

Although the introduction of an angular dependence through
eq.~(\ref{FACTOR}) provides a quite adequate description of all
the distributions, with no need for the $S$--wave FSI, it seems
unlikely that this would be suppressed completely. To test some of
the possibilities, we have adjusted the contribution of the FSI to
provide a slightly better description of the data; the results
shown by dotted lines in Fig.~\ref{dNwdVar} were achieved by
decreasing the FSI enhancement by factors of $0.8$ and $0.4$ at $Q
=60\,$MeV and 92\,MeV, respectively. The values of the total cross
sections obtained on the basis of these various assumptions are
presented in Table~\ref{xvalues}. Other variants, including
allowing some dependence on $\Theta_{pp}$, invariably give results
that lie within the ranges shown in the table.

\begin{table}
\caption{Total cross sections for different models. The form of
the FSI factor is given in eq.~(\ref{FSI}) and that of the assumed
angular dependence in eq.~(\ref{FACTOR}).} \label{xvalues}
\begin{center}
\begin{tabular}{lcc}
\hline\noalign{\smallskip}
 & $Q=60\,$MeV & $Q=92\,$MeV \\
\noalign{\smallskip}\hline\noalign{\smallskip}
Model & $a = 0.2$ & $a = 0.6$ \\
 & $k = 0.8$ & $k = 0.4$ \\
\noalign{\smallskip}\hline\noalign{\smallskip}
PS & 5.0~$\mu$b & 10.8~$\mu$b \\
PS\,+\,FSI(100\%) & 7.1~$\mu$b & 14.3~$\mu$b \\
PS\,+\,$a\, P_2(\cos\Psi_{pp})$ & 4.8~$\mu$b & 10.4~$\,\mu$b \\
PS\,+\,$k\times$FSI\,+\,$a\,P_2(\cos\Psi_{pp})$ & 6.6~$\mu$b & 12.0~$\mu$b \\
\noalign{\smallskip}\hline
\end{tabular}
\end{center}
\end{table}

\newpage
\section{Results and Discussion}
\label{xtot}

Combining the results of all the variants that can hardly be
distinguished within the errors, we find total cross sections
lying in the ranges $4.8-7.1\mu$b at $Q=60\,$MeV and 
$10.4-14.3\mu$b at 92\,MeV, respectively.  The highest values are
achieved with the rather unrealistic pure FSI model and so we take
rather the last two lines of Table~\ref{xvalues} as an indication
of the model--dependence of the results. These give
\begin{tabbing}
\hspace*{0.2cm}\=$\sigma(Q=60\,$MeV$) =$\= $\,\phantom{1}
(5.7 \pm 0.6_{\rm stat} \pm 0.8_{\rm syst} \pm 0.9_{\rm mod})\:\mu$b,\\
\>$\sigma(Q=92\,$MeV$) =$\> $\,(11.2 \pm 1.1_{\rm stat} \pm
1.7_{\rm syst} \pm 0.8_{\rm mod})\:\mu$b\,,
\end{tabbing}
where the first error represents the statistical uncertainty. The
second figure stands for the total experimental systematic
uncertainty, which includes contributions from the luminosity
determination, the possible change of acceptance due to
uncertainties in the detector positioning, but mainly the
determination of the total number of $pp\omega$ events. These were
all treated as independent errors. The final number is our estimate of
the influence of the model on the acceptance calculation.

\begin{figure}[htb]
\begin{center}
\resizebox{0.45\textwidth}{!}{%
 \includegraphics{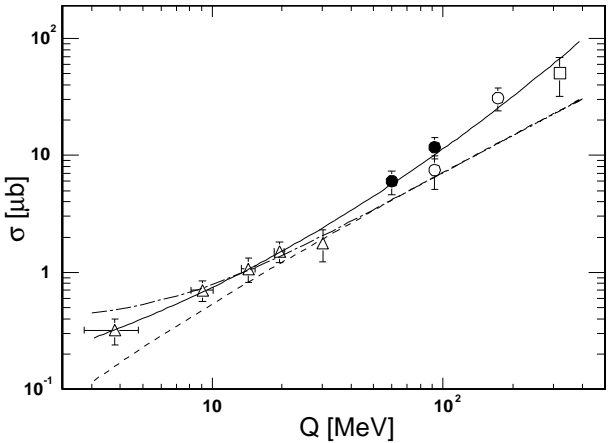}}
\end{center}
\caption {The energy dependence of the $pp\rightarrow{}pp\omega$
total cross section. Our two points (closed circles) are compared
with earlier data from Refs.~\cite{SPES} (triangles), \cite{DISTO}
(square), and \cite{TOF} (open circles). In all cases, systematic
and statistical eros have been added in quadrature. The chain and
dashed curves show respectively the normalised energy dependence
of eq.~(\ref{1}) with and without smearing over the finite
$\omega$ width. The solid line is a purely phenomenological
interpolation of the whole data set given by
eq.~(\ref{fitting}).\label{Xsec_tot}}
\end{figure}

Our $pp\to{}pp\omega$ total cross sections are shown in
Fig.~\ref{Xsec_tot} with vertical error bars estimated from the
quadratic sum of both the systematic and statistical errors
presented above. Also shown are the results of other experiments
at low $Q$~\cite{SPES,DISTO,TOF}. The 92\,MeV value is slightly
higher than that of TOF, though the error bars do overlap. With
the possible exception of the highest SPESIII point, the results
are consistent with a smooth energy dependence and a global fit to
the whole data set yields
\begin{equation}
\label{fitting} \log_{10}\sigma =
-0.855+0.495\log_{10}Q+0.230\left(\log_{10}Q\right)^2,
\end{equation}
where $Q$ is measured in MeV and the cross section in $\mu$b. This
is also shown in the figure.

In the SPESIII work~\cite{SPES}, the energy dependence of the
total cross section was compared with a simplistic model,
consisting only of three--body phase space but modified by the
$S$--wave $pp$ FSI~\cite{GFW}. This leads to the analytic result:
\begin{equation}
\label{1} \sigma(pp\to pp\omega) =
C_{\omega}\:\left(\frac{Q/\epsilon}
{1+\sqrt{1+Q/\epsilon}}\right)^{\!\!2},
\end{equation}
where $\epsilon\approx 0.45$~MeV. However, at low excess energies
it is important to smear this function over the width of the
$\omega$--meson. Both these forms are shown in
Fig.~\ref{Xsec_tot}, with the value of $C_{\omega}=37\,$nb being
chosen to fit the SPESIII points rather than those at higher
energy. Although the $S$--wave assumption is clearly untenable at
larger $Q$, the curves do suggest that $P$ or higher partial waves are
important there and a good representation of the data can be
obtained by adding a small amount of pure phase--space
distribution to the FSI of eq.~(\ref{1}). While presenting a
baseline against which one can judge the more refined theoretical
models, this approach also shows that the finite--width effects
are negligible in our energy domain.

\begin{figure}[htb]
\begin{center}
\resizebox{0.45\textwidth}{!}{%
 \includegraphics{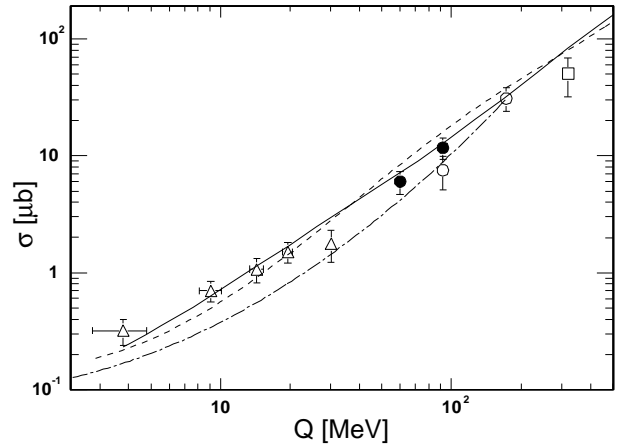}}
\end{center}
\caption {Experimental data on the $pp\rightarrow{}pp\omega$ total
cross section shown in Fig.~\ref{Xsec_tot} compared with the
predictions from microscopic theoretical models shown by the
chain (\cite{Nakayama}), solid (\cite{Kampfer}), and
dashed (\cite{Fuchs}) curves. \label{Xsec_tot2}}
\end{figure}

The predictions of three microscopic
models~\cite{Nakayama,Kampfer,Fuchs} are shown in
Fig.~\ref{Xsec_tot2}. All of them are versions of a
meson--exchange model, but with different emphasis on the role of
the $N^*$ isobars. It should be noted that these calculations
involve a large number of parameters, some of which have
been adjusted in order to optimise agreement with
the published data, including the differential distribution at
173\,MeV~\cite{TOF}.

In the early work of Tsushima and Nakayama~\cite{Nakayama},
nucleonic and mesonic current contributions were considered but
the energy dependence of the total cross section is described
better by the inclusion of effects of nucleon resonances, and it
is that curve which is shown in Fig.~\ref{Xsec_tot2}.
Nevertheless, this calculation still underestimates the SPESIII
data by about a factor of two even though the finite $\omega$
width was taken into account. Despite neglecting this, and the
initial state interaction, Kaptari and K\"ampfer~\cite{Kampfer}
got good agreement with the then available data without relying on
any $N^*$ contribution. It was shown there that the effects of the
$S$--wave FSI are particularly important, for without it a
phase--space energy dependence of the $Q^2$ type is predicted. The
T\"{u}bingen group~\cite{Fuchs} had extensive recourse to several
nucleon resonances, especially the $N^*(1535)$, but the novel
feature of their approach is the coupling to off--shell
$\omega$--mesons, which is particularly important close to the
reaction threshold. If this is correct then one would expect
interesting effects when the $\omega$ is measured exclusively
through its $e^+e^-$ decay mode~\cite{Fuchs}.

Having established the energy dependence of the total cross
section for $pp\to{}pp\omega$, we are now in a position to
investigate further the OZI rule in near--threshold production
processes. If the $(\omega,\phi)$ mixing in the vector meson nonet
were ideal, such that the $\phi$ were a pure $s\bar{s}$ state,
then its three-pion decay would be forbidden by the OZI
rule~\cite{OZI}. That it takes place at all is interpreted as an
indication that the mixing is not quite ideal. Using the
Gell-Mann--Okubo mass formulae, Lipkin~\cite{Lipkin} estimated
that the ratio of coupling strengths of the $\phi$ and $\omega$ to
non--strange hadrons should be
\begin{equation}
\label{ozi_lipkin}%
R_\mathrm{OZI}=4.2\,\times 10^{-3}\:.
\end{equation}
It is therefore useful to evaluate the ratio of the $\phi$ and
$\omega$ production cross sections in $pp$ collisions at the same
value of the excess energy:
\begin{equation}
R_{\phi/\omega}\equiv\frac{\sigma(pp\to{}pp\phi)}{\sigma(pp\to
pp\omega)}\,\cdot
\end{equation}

Taking the recent measurement~\cite{Hartmann} of the total cross
section for $\phi$ production in $pp$ collisions at $Q=18.5$,
34.5, and 75.9\,MeV in conjunction with the global fit of
eq.~(\ref{fitting}) to the data set of Fig.~\ref{Xsec_tot}, we
find at these energies $R_{\phi/\omega}=(3.1\pm0.6)\times
10^{-2}$, $(3.0\pm0.7)\times 10^{-2}$, and $(2.4\pm0.7)\times
10^{-2}$, respectively, where all error bars have been compounded
quadratically. Applying the same procedure to the $\phi$ point
reported by the DISTO collaboration leads to
$R_{\phi/\omega}=(2.2\pm1.0)\times 10^{-2}$ at 83\,MeV. The mean
value of $R_{\phi/\omega}=(2.8\pm0.4)\times 10^{-2}\approx 7\times
R_\mathrm{OZI}$, is lower than that obtained in
Ref.~\cite{Hartmann} because of the higher values for $\omega$
production that are now apparent in Fig.~\ref{Xsec_tot}. Although
not statistically significant, there are indications from these
numbers that $R_{\phi/\omega}$ might decrease with rising $Q$ but,
to test this, data would be needed on $\phi$ production at higher
values of $Q$. However, it should be noted that such a slow
decline had been predicted in model calculations. This arises
there through the decreasing effect of the destructive
interference between the nucleonic and mesonic contributions to
$\omega$ production as the energy is raised~\cite{Speth}.

\section{Conclusions}
\label{conclude}

We have presented new measurements of the $pp\to{}pp\omega$
reaction at excess energies of 60 and 92\,MeV. The SPESIII
technique~\cite{SPES} of kinematically shifting the data at one
energy to estimate the background at a neighbouring energy was
successfully employed for differential as well as total spectra.
In this way the small $\omega$ signal in the missing--mass
distribution could be identified despite the large amount of
multipion production. With the setup actually employed for
studying this reaction at ANKE, the acceptance, especially at
small $\varepsilon_{pp}$, was limited. A variety of assumptions
were tested against one--dimensional spectra in order to assess
the model--dependence of the total $\omega$ production cross
section that we obtained in this way. This led to uncertainties
that were comparable to the statistical and other systematic
errors. The value for the total cross section at 92\,MeV is a
little higher than that found by the TOF group~\cite{TOF}, but our
two points join smoothly with the results of other experiments.
There is evidence from the anisotropic $pp$ angular distribution
at 92\,MeV that $P$ or higher $pp$ partial waves are significant at
this energy.

The new data allow us to extract the OZI ratio with greater
confidence and smaller error bars than before. This results in a
slightly smaller value of $R_{\phi/\omega}$ than previously
quoted~\cite{Hartmann}, with possibly a hint of some energy
dependence  (but see also Ref.~\cite{Sibirtsev}).

All three theoretical calculations
considered~\cite{Nakayama,Kampfer,Fuchs} appeared after the broad
outlines of the energy dependence of the total cross section had
been determined experimentally and in some cases the model
parameters were adjusted to reproduce this. Other experimental
observables such as differential cross sections or decay angular
distributions are needed to constrain the models more tightly and
it is encouraging to note that analysing power results will soon be
made available by the COSY--TOF collaboration~\cite{Brinkmann}.
Since good data exist on $pn\to d\phi$ at low energies~\cite{Yoshi},
further data on $\omega$ production with a neutron target would
be particularly helpful~\cite{pndomega}.

The WASA at COSY facility will soon become
operational~\cite{WASA}. The production of the $\omega$ in $pp$
collisions could then be investigated, with larger geometrical
acceptance and lower background, \emph{via} the detection of the
$\omega \to \pi^0 \gamma$ decay ($BR\approx9\%$) and the
reconstruction of its invariant mass. Such an approach would also
have the advantage of leading to a determination of the tensor
polarisation of the $\omega$, a quantity which is very sensitive
to the presence of higher partial waves. A similar sensitivity
also exists in the spin--spin correlation in
$\stackrel{\to}{p}\stackrel{\to}{p}\to pp\omega$,
which could be measured at ANKE~\cite{SPIN}.

\begin{acknowledgement}
Useful discussions with K.\,Nakayama and other members of
the ANKE Collaboration are gratefully acknowledged. Our thanks
extend also to the COSY machine crew for their support as well as
to B.\,Zalikhanov for his help with the MWPCs. This work has
partially been supported by the BMBF, DFG, Russian Academy of
Sciences, and COSY FFE.
\end{acknowledgement}

\end{document}